\documentclass[
 amsmath,amssymb,
 floatfix, 
 reprint,
 superscriptaddress,
 pra,
]{revtex4-2}

\usepackage{amsmath}
\usepackage{xcolor}
\usepackage{graphicx}
\usepackage{bbold}
\graphicspath{ {./} }
\usepackage{dcolumn}
\usepackage{bm}
\usepackage{float}

\newcommand{\iitm}{\affiliation{Department of Physics, Indian Institute of Technology Madras, Chennai 600036, India}}

\newcommand{\cquicc}{\affiliation{ Center for Quantum Information, Communication and Computing, Indian Institute of Technology Madras, Chennai 600036, India}}

\begin{document}
\raggedbottom

\newcommand{\bra}[1]{\langle #1 |}
\newcommand{\ket}[1]{| #1 \rangle}
\newcommand{\braket}[2]{\langle #1 | #2 \rangle}
\newcommand{\exb}{\mathbf{E} \times \mathbf{B}}
\newcommand{\numberthis}{\addtocounter{equation}{1}\tag{\theequation}}
\title{Adiabatic Cooling of Planar Motion in a Penning Trap Ion Crystal to Sub-Millikelvin Temperatures}

\author{Wes Johnson}
\email{Wes.Johnson@colorado.edu}
\affiliation{Department of Physics, University of Colorado, Boulder, Colorado 80309, USA}
\author{Bryce Bullock}
\affiliation{National Institute of Standards and Technology Boulder}
\affiliation{Department of Physics, University of Colorado, Boulder, Colorado 80309, USA}
\author{Athreya Shankar}
\iitm
\cquicc
\author{John Zaris} 
\affiliation{Department of Physics, University of Colorado, Boulder, Colorado 80309, USA}
\author{John J. Bollinger}
\affiliation{National Institute of Standards and Technology Boulder}
\author{Scott E. Parker}
\altaffiliation[Also ]{Renewable and Sustainable Energy Institute, University of Colorado, Boulder}
\affiliation{Department of Physics, University of Colorado, Boulder, Colorado 80309, USA}

\date{\today}

\begin{abstract} 
Two-dimensional planar ion crystals in a Penning trap are a platform for quantum information science experiments. 
However, the low-frequency planar modes of these crystals are not efficiently cooled by laser cooling, which can limit the utility of the drumhead modes for quantum information processing. 
Recently, it has been shown that nonlinear mode coupling can enhance the cooling of the low-frequency planar modes.
Here, we demonstrate in numerical simulations that this coupling can be dynamically tuned by adiabatically changing the rotation frequency of the ion crystal during experiments. 
Furthermore, we show that this technique can, in addition, produce lower temperatures for the low-frequency planar modes via an adiabatic cooling process.
This result allows cooling of the planar modes to sub-millikelvin temperatures, resulting in improved spectral resolution of the drumhead modes at experimentally relevant rotation frequencies, which is crucial for quantum information processing applications.
\end{abstract}

\maketitle

\section{\label{sec:intro}Introduction}

Single-plane (two-dimensional) ion crystals in a Penning trap provide a robust platform for quantum science experiments with hundreds of ions~\cite{Monroe2021}.
In these systems, laser interactions couple the internal states of the ions via their out-of-plane vibrational modes, known as drumhead modes, which serve as the quantum buses for mediating interactions.    
Two-dimensional ion crystals have been used to simulate many-body quantum dynamics~\cite{Britton2012, Bohnet2016, Shankar2020} and perform quantum-enhanced sensing~\cite{Gilmore2017, Affolter2020, Gilmore2021}. 
Experiments have also demonstrated single-site addressability \cite{McMahon2024}, and site-resolved readout and detection \cite{Wolf2024}.
Recent theoretical work has proposed extending these systems to three-dimensional configurations \cite{Hawaldar2024}, but to date, experimental demonstrations have been limited to two-dimensional crystals.  

However, the in-plane (planar) modes of the ion crystal—particularly the low-frequency $\exb$ modes—are not efficiently cooled by standard Doppler cooling techniques. 
This presents a challenge for experiments, since residual energy in these modes broadens the drumhead mode spectrum \cite{Shankar2020}, limiting spectral resolution and the ability to perform high-fidelity quantum operations.    
Recent work~\cite{Johnson2024} showed that nonlinear coupling between motional branches can enable enhanced cooling of the planar modes, but only under specific conditions.

Here we demonstrate that this nonlinear coupling can be dynamically tuned by adiabatically varying the ion crystal’s rotation frequency. 
We show in simulation that this approach enables \textbf{adiabatic cooling} of the planar modes to sub-millikelvin temperatures --- without relying on additional laser beams or significantly changing the experimental setup. 
Remarkably, this effect emerges from a simple ramp in rotation frequency, provided the ramp in frequency is sufficiently slow and smooth.

We begin by reviewing the Penning trap setup and the role of the rotating wall potential in controlling the crystal’s rotation in Section~\ref{sec:setup}.
We then introduce the theoretical framework for adiabatic cooling, including a simple energy-scaling model based on action conservation in Section~\ref{sec:adiabaticCooling}.
Finally, we present simulations that validate this model across a range of scenarios—from idealized initial conditions to realistic thermal ensembles—and show that adiabatic ramps can be combined with laser cooling to achieve well-resolved mode spectra in experimental settings in Section~ \ref{sec:results}. 

\section{\label{sec:setup}Background and Setup}

To make our work concrete, we consider the National Institute of Standards and Technology (NIST) Penning trap experimental setup~\cite{Bohnet2016}.  
The laser cooling simulation parameters and setup are described in Ref.~\cite{Johnson2024} and values used here given in the Appendix~\ref{app:simulations}. 

In the Penning trap, the ion crystal rotates in the $xy$-plane with a rigid-body rotation frequency, $\omega_r$, about the $z$-axis.
The ion crystal rotation is phase-locked to an applied rotating wall potential at a frequency $\omega_r$. 
In a frame co-rotating with the ion crystal, the potential energy of the ions is given by \cite{Tang2019}: 
\begin{equation}
\begin{split}
U\left(\{\mathbf{R}_i\}_{i=1}^N\right) = \frac{1}{2} \sum_{i=1}^{N}\sum_{j \ne i} \frac{k_e q^2}{|\mathbf{R}_i - \mathbf{R}_j|} \\
+\sum_{i=1}^N \frac{1}{2} m \omega_z^2 \left( z_i^2
+ \left( \beta + \delta \right) x_i^2 + \left( \beta - \delta \right) y_i^2 \right).
\end{split}
\label{eqn:potentialEnergy}
\end{equation}
Here, $\mathbf{R}_i = (x_i, y_i, z_i)$ is the position vector of the $i$th ion in the co-rotating frame. 
The parameters $m$ and $q$ are the mass and charge of a $^9\text{Be}^+$ ion, respectively, and $k_e$ is the Coulomb constant.  
The rotating wall strength $\delta$ controls the radial anisotropy of the ion crystal. 
Finally, $\omega_z$ is the axial confinement frequency (parallel to the magnetic field $\mathbf{B}$), which determines the oscillation frequency of a single ion along the $z$-axis.
The planar confinement parameter $\beta$ defines the relative strength of the planar to axial confinement, and is given by:
\begin{equation}
\beta = \frac{\omega_r\left(\omega_c-\omega_r\right)}{\omega_z^2} - \frac{1}{2}, 
\label{eqn:beta}
\end{equation}
where $\omega_c$ is the cyclotron frequency of the ions.

In this work, we explore how dynamically changing the rotation frequency, $\omega_r$, of the ion crystal during experiments can enhance the cooling of the planar motions. 
Specifically, we simulate the dynamics of the ion crystal under a controlled ramp in the frequency of the rotating wall potential from an initial frequency $\omega_i$ to a final frequency $\omega_f$ with a ramp time $T$.    

To avoid impulsive forces that arise from discontinuities in the derivative of the rotating wall frequency, we use a half-cosine ramp to smoothly turn on and off the ramp:
\begin{equation}
\omega_r(t) = \frac{\omega_i + \omega_f}{2} + \frac{\omega_i - \omega_f}{2}\cos\left(\frac{\pi t}{T}\right).
\label{eqn:cosineRamp}
\end{equation}
The impact of different ramping protocols on the ion crystal is discussed in Section~\ref{sec:adiabaticCooling}.    
We find that the half-cosine ramp avoids heating and retains adiabaticity. 
Other smooth ramps with zero derivatives were investigated and yielded similar results; however, the half-cosine ramp was chosen for its simplicity. 
Therefore, we use the half-cosine ramp in all simulations except where otherwise noted.

\section{\label{sec:adiabaticCooling}Adiabatic Cooling of Planar Motion}

Under NIST experimental parameters \cite{Gilmore2021}, the normal modes of the ion crystal separate into three distinct branches \cite{Wang2013, Dubin2020, Shankar2020}: the low-frequency $\exb$ modes ($\approx$ 1 - 100 kHz), the intermediate frequency drumhead modes ($\approx 1$ MHz), and the high-frequency cyclotron modes ($\approx 10$ MHz). 
Importantly, in contrast to the cyclotron modes---which due to their kinetic nature couple strongly to the in-plane Doppler laser cooling beams---the $\exb$ modes are dominated by potential energy contributions and are not efficiently cooled by the laser cooling beams \cite{Johnson2024}.  
Physically, the slow in-plane motion is magnetron-like in the rotating frame: a single ion executes nearly circular drift at an almost fixed radius, so the planar potential energy remains large while the planar kinetic energy is small, meanwhile the fast cyclotron-like motion is dominated by kinetic energy. 
Following Ref.~\cite{Shankar2020}, we quantify this unequal partitioning \emph{per mode $n$} by $R_n \equiv \overline{\mathrm{PE}}_n/\overline{\mathrm{KE}}_n$.
Assuming only a single ion in the trap, in a circular orbit of radius r, the time-averaged planar potential energy is $\overline{\mathrm{PE}}=(1/2)m\omega_\perp^2 r^2 = m\beta\omega_z^2 r^2/4$, while $\overline{\mathrm{KE}}=(1/2)m\omega_\pm^2 r^2$ for the cyclotron ($+$) and magnetron ($-$) modes.
Using the single-ion identity $\omega_+\omega_-=\omega_\perp^2=\beta\omega_z^2/2$, one finds $R_+ = \omega_-/\omega_+ \ll 1$ and $R_- = \omega_+/\omega_- \gg 1$. 
More generally, in the case of a multi-ion crystal, the 2N planar modes split into a low-frequency ($\exb$) and a high-frequency (cyclotron) branch of $N$ modes each, which are respectively dominated by $\mathrm{PE}$ and $\mathrm{KE}$, with $R_n\gg 1$ for the slow $\exb$ modes and $R_n \ll 1$ for the fast cyclotron modes~\cite{Shankar2020}.

Due to the large value of $R_n$ for the $\exb$ modes, cooling of the crystal's $\mathrm{PE}$ is not efficient under typical experimental conditions.
Past work suggests that the low-frequency $\exb$ modes are likely not cooled below 10 mK for the NIST Penning trap setup~\cite{Shankar2020}, which employs a strong 4.5 T magnetic field.
In previous work \cite{Johnson2024}, we showed that the low-frequency $\exb$ modes can be rapidly cooled to millikelvin temperatures using a near-resonant nonlinear coupling to the out-of-plane (drumhead) modes. 
However, this required employing higher than usual rotation frequencies ($\omega_r$) of the ion crystal to generate low-frequency drumhead modes that provide the near-resonant nonlinear coupling. 
However, for quantum information processing applications, which typically require small Lamb-Dicke parameters \cite{Wineland1998,Leibfried2003} of the drumhead modes, the rotation frequency $\omega_r$ of the ion crystal is typically lowered, moving the drumhead modes out of resonance with the $\exb$ modes.  
Following NIST Penning-trap conventions, the relevant quantity is the \emph{per-ion} LD parameter for the addressed transition, which depends on the laser geometry and a sum over drumhead-mode participations~\cite{Britton2012,Bohnet2016}.
Its explicit form is given in the supplemental materials of Refs.~\cite{Britton2012,Bohnet2016}; here we emphasize the key scaling that for fixed geometry $\eta \propto \omega_m^{-1/2}$, so softer modes contribute more. 
In our parameter regime, lowering $\omega_r$ raises the drumhead-mode frequencies and thus reduces the effective LD parameter, aiding LD regime operation.
However, lowering $\omega_r$ also removes the $\exb$-to-drumhead cooling mechanism present at higher rotation frequency, motivating our adiabatic ramp protocol.
Therefore, we demonstrate a method to dynamically tune the rotation frequency of the ion crystal during experiments to enhance the cooling of the $\exb$ modes while operating at lower $\omega_r$.

To understand how varying the rotating wall frequency affects the normal mode energies, we express the Hamiltonian for each normal mode in the action-angle action-variable canonical form. 
A detailed derivation of the normal mode decomposition in canonical coordinates is given in Appendix~\ref{app:normalModes}. 
Each normal mode can be described by a harmonic oscillator Hamiltonian \cite{Dubin2020}, with the following energy: 
\begin{equation}
    E = \frac{1}{2}\omega (Q^2 + P^2) = \omega I,
\label{eqn:mode_hamiltonian}
\end{equation}
where $Q$ and $P$ are the generalized canonical position and momentum variables of the mode, respectively, $\omega$ is the frequency of the mode, and $I$ is the action variable, an \textbf{adiabatic invariant}.   
According to the adiabatic invariance of the action variable for a harmonic oscillator, the ratio $E/\omega$ is constant if the frequency of the mode is varied on a time scale much slower than the oscillator period \cite{LandauMechanics1976}.
This yields the following relationship between the energy of the mode after the adiabatic variation of the mode's frequency:
\begin{equation}
    E_f = E_i \left(\frac{\omega_f}{\omega_i}\right),
\label{eqn:adiabaticEnergy}
\end{equation}
where $E_i$ and $E_f$ are the energy of the mode before and after the adiabatic variation of the mode's frequency.

In Appendix \ref{app:adiabaticEnergy}, we show that the frequencies of the $\exb$ modes are approximately proportional to the planar confinement parameter in Eq.~(\ref{eqn:beta}), $\beta$, which is in turn related to the rotating wall frequency, $\omega_r$.   
This yields a simple expression for the energy of \emph{each} $\exb$ mode before and after the ramp:
\begin{equation}
    E_f^{\exb} = E_i^{\exb} \left(\frac{\beta_f}{\beta_i}\right),
\label{eqn:exbAdiabaticEnergy}
\end{equation}
where $\beta_i$ and $\beta_f$ are the initial and final planar confinement parameters and $E_i^{\exb}$ is the initial energy of the $\exb$ mode.
Hence, if the rotation frequency of the ion crystal is decreased during the ramp, the energy of the $\exb$ modes is reduced.
This suggests that the $\exb$ modes can be cooled further by lowering the rotation frequency of the ion crystal, preparing them in a lower energy state than what is accessible with laser cooling alone at typical rotation frequencies, $\omega_r$.

To ensure that the ion crystal remains in a fixed, stable equilibrium configuration during the ramp, and hence that the mode frequencies change smoothly, all of the following conditions must be satisfied:  

1. \textbf{Fixed $\delta/\beta$ Ratio:} The ratio of the rotating wall strength $\delta$ to the planar confinement parameter $\beta$ must be fixed during the ramp.
Otherwise, if $\beta$ decreases while $\delta$ remains constant, the ion crystal becomes more elongated in the $y$-direction, leading to deformation and heating. 
This effect is illustrated in Section~\ref{sec:results} in Fig.~\ref{fig:compareRamps}(a). 

2. \textbf{Smooth ramping:} The ramp must be smoothly turned on and off to prevent heating of the collective rocking motion of the ion crystal due to the impulsive Euler force \cite{LandauMechanics1976}.
A half-cosine ramp avoids discontinuities in $\dot{\omega}_r$, and hence suppresses the Euler force, and mitigates the heating of the rocking mode during the ramp. 
This is demonstrated in Section~\ref{sec:results} in Fig.~\ref{fig:compareRamps}(c).    

3. \textbf{Adiabatic ramp time:} The ramp time must be sufficiently long compared to the periods of the lowest frequency modes of the ion crystal, particularly the $\exb$ modes, to maintain adiabatic evolution. 
See Fig.~\ref{fig:varyWallStrength}(a) and Fig.~\ref{fig:varyWallStrength}(b).

The above three criteria are practical \emph{sufficient} conditions for adiabatic following of all $3N$ normal modes during the ramp and for preserving the crystal's equilibrium configuration up to an overall rescaling. 
They are motivated, respectively, by (i) avoiding changes to the aspect ratio via a fixed $\delta/\beta$, (ii) suppressing impulsive excitation from the Euler force by ensuring $\dot{\omega}_r$ is continuous (e.g., a half-cosine ramp), and (iii) choosing a ramp duration long compared to the slowest mode periods so that $|\dot{\omega}_n|/\omega_n^2 \ll 1$ throughout. 
These are not necessary conditions—other ramping schemes may also succeed; for example, single-ion shortcuts and related non-adiabatic protocols in time-dependent magnetic field Penning traps provide alternative routes to target states~\cite{Kiely2015}.

\section{\label{sec:results}Results}

\begin{figure}
\includegraphics[width=\linewidth]{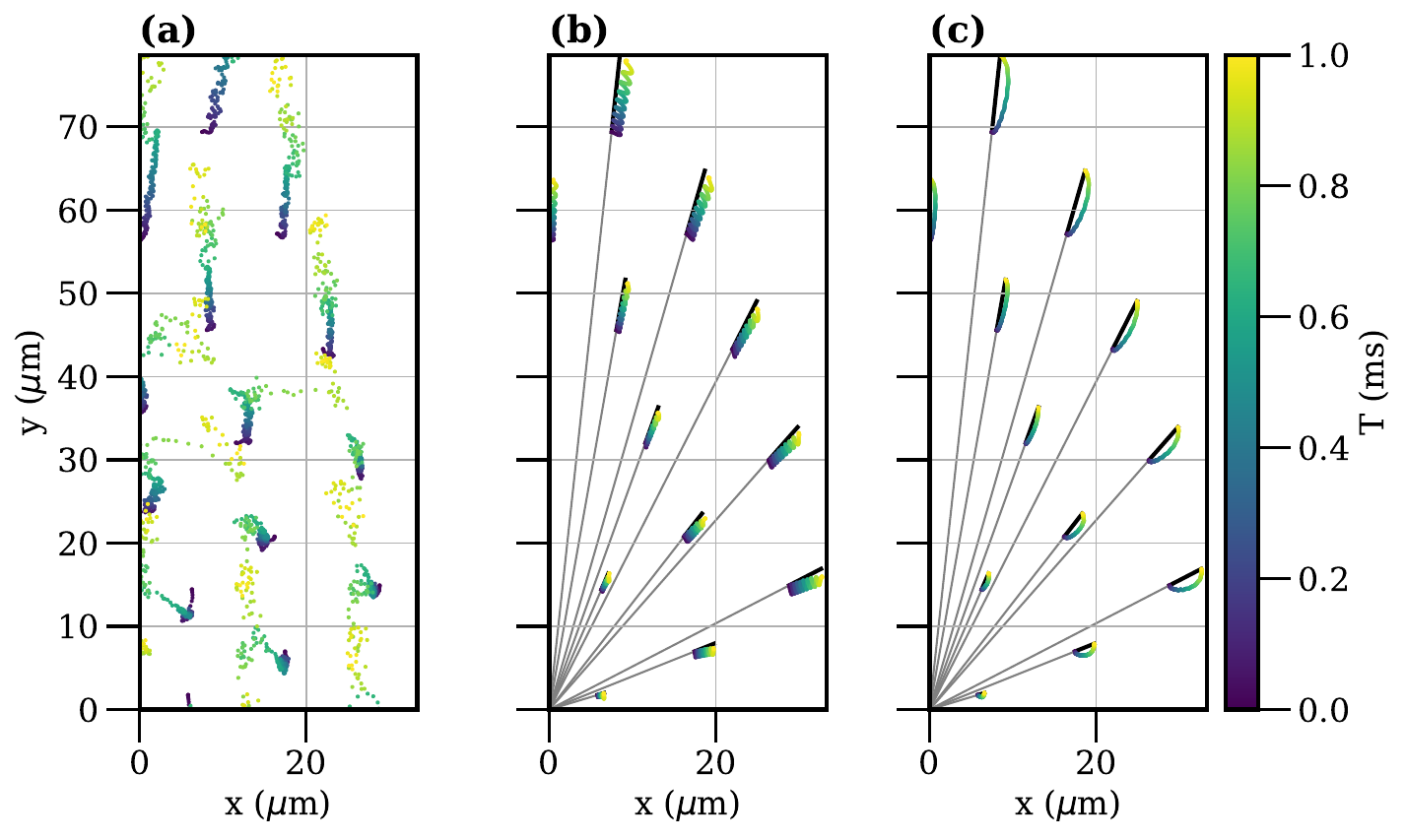}
\caption{\label{fig:compareRamps}
\textbf{Comparison of linear and half-cosine rotating wall frequency ramps.}
A crystal of $N = 50$ ions is initialized in the zero-temperature equilibrium configuration, and the rotation frequency is ramped from 200 to 190 kHz over 1 ms using three protocols: \textbf{(a)} a linear ramp with $\delta$ held constant while $\beta$ is lowered, \textbf{(b)} a linear ramp with fixed $\delta/\beta$, and \textbf{(c)} a half-cosine ramp with fixed $\delta/\beta$.
Each panel shows the ion trajectories in the first quadrant of the $xy$-plane during the ramp; color indicates time.
Without fixing $\delta/\beta$ [\textbf{(a)}], the crystal elongates and reorganizes.
Fixing $\delta/\beta$ [\textbf{(b)}] prevents reconfiguration but excites a rocking mode due to abrupt ramp edges.
The half-cosine ramp [\textbf{(c)}] preserves crystal structure and avoids rocking motion.
\textbf{(b,c)} Light-gray segments from each ion to the origin indicate global rescaling (collinearity with preserved aspect ratio), and solid black segments connect each ion's initial equilibrium site to its final rescaled equilibrium site from Eq.~(\ref{eq:rescaled_eq}), tracing the instantaneous-equilibrium path.
For the short half-cosine ramp \textbf{(c)} ($T=1~\mathrm{ms}\sim 2\pi/\omega_{\min}$) Euler-force excitation causes brief departures from these paths, but ions return and come to rest at $t=T$, whereas the equal-duration linear ramp \textbf{(b)} leaves residual azimuthal rocking.
}       
\end{figure}

\begin{figure} 
\includegraphics[width=\linewidth]{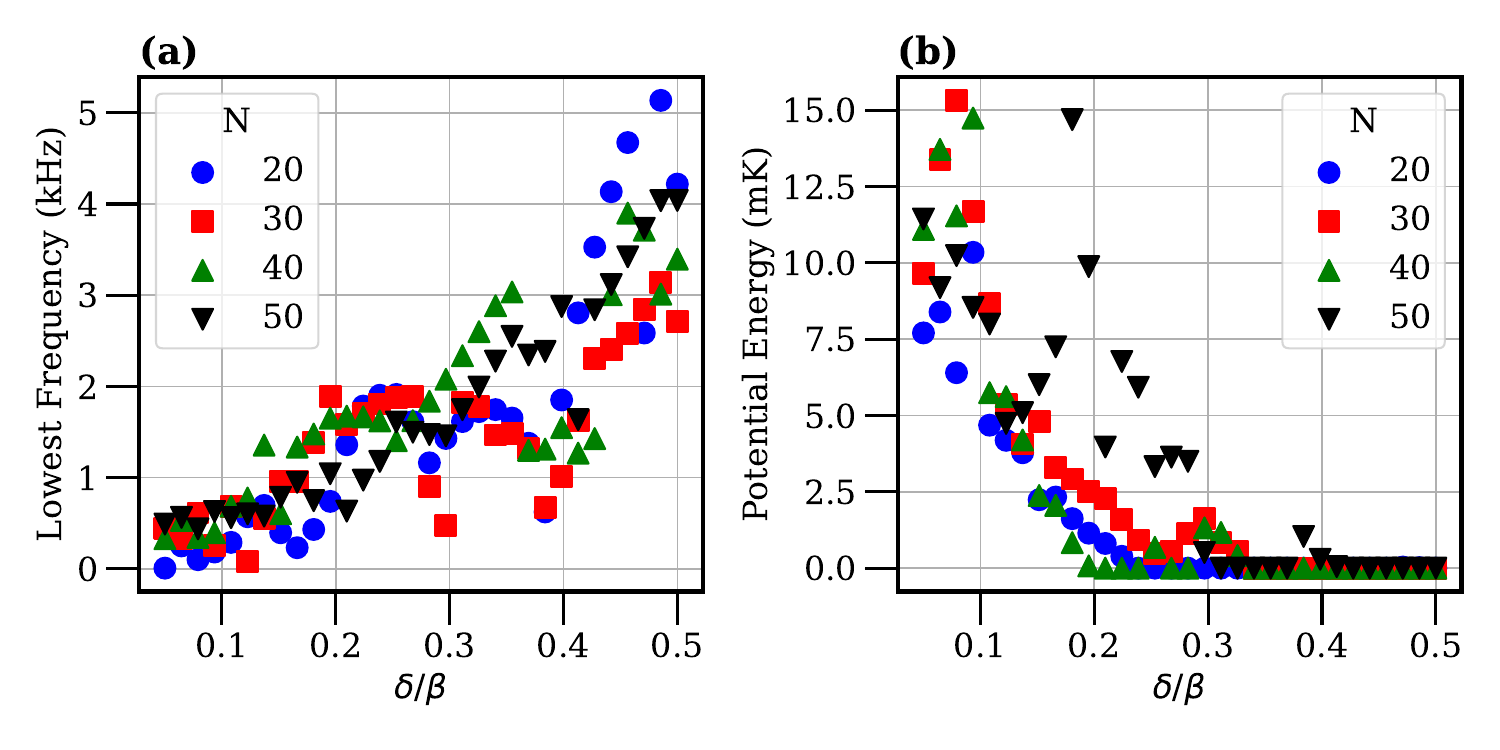}
\caption{\label{fig:varyWallStrength}
\textbf{Effect of rotating wall strength on lowest mode frequencies and crystal stability.}
Ion crystals with $N = 20$, $30$, $40$, and $50$ ions are initialized in the minimum potential energy (zero temperature) configuration at $\omega_r = 2\pi \times 200$ kHz with varying rotating wall strengths in the range $\delta/\beta = 0.1$ to $0.5$. 
Each system is then ramped to 180 kHz over 1 ms using a half-cosine ramp.
\textbf{(a)} The frequency of the lowest $\exb$ mode increases with rotating wall strength, indicating improved adiabatic conditions at larger $\delta/\beta$.
\textbf{(b)} The average potential energy offset from $E = 0$ equilibrium, measured over 1 ms following the ramp, decreases with increasing $\delta/\beta$. 
For $\delta/\beta \gtrsim 0.3$, the system remains near the $E = 0$ equilibrium, indicating minimal heating during the ramp.
}
\end{figure}

\begin{figure}
\includegraphics[width=\linewidth]{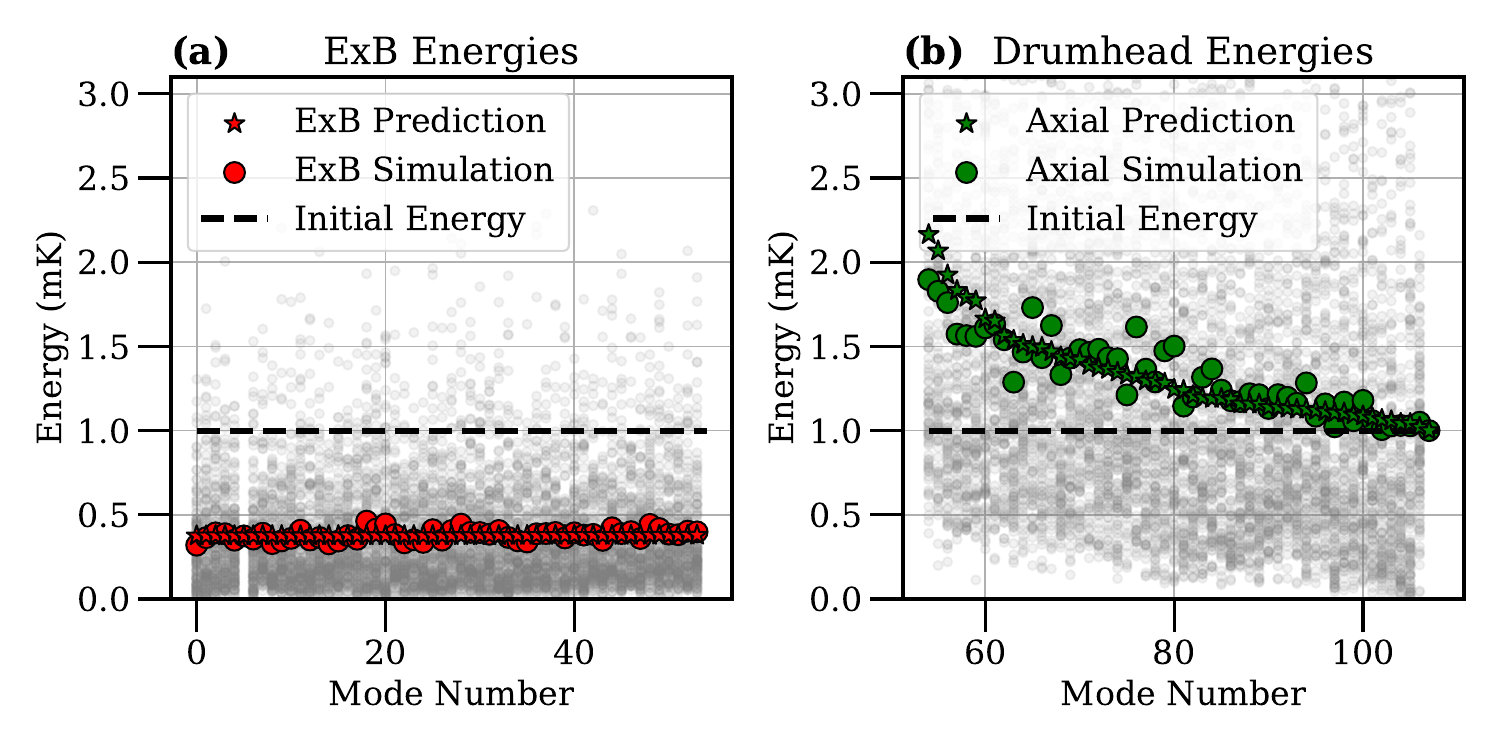}
\caption{\label{fig:adiabaticCooling}
\textbf{Adiabatic cooling of planar motion compared to theory.}
An ensemble of 128 different initializations of a 54-ion crystal is generated with all mode amplitudes corresponding to a temperature of 1 mK. 
A half-cosine ramp is applied to decrease the rotating wall frequency from 200 kHz to 180 kHz over 20 ms.
\textbf{(a)} Final energies of the $\exb$ modes are plotted versus mode number.
Small gray dots (128 per mode) represent different ion crystal initializations, and large red dots represent the ensemble averages.
The average energies are reduced to approximately 0.4 mK, consistent with the adiabatic cooling prediction (red stars). 
This confirms that the adiabatic energy scaling model accurately describes the cooling process for the $\exb$ modes.
\textbf{(b)} Final energies of the axial (drumhead) modes. 
Small gray dots (128 per mode) represent different ion crystal initializatons, and large green dots represent the ensemble averages.
}
\end{figure}

\begin{figure}
\includegraphics[width=\linewidth]{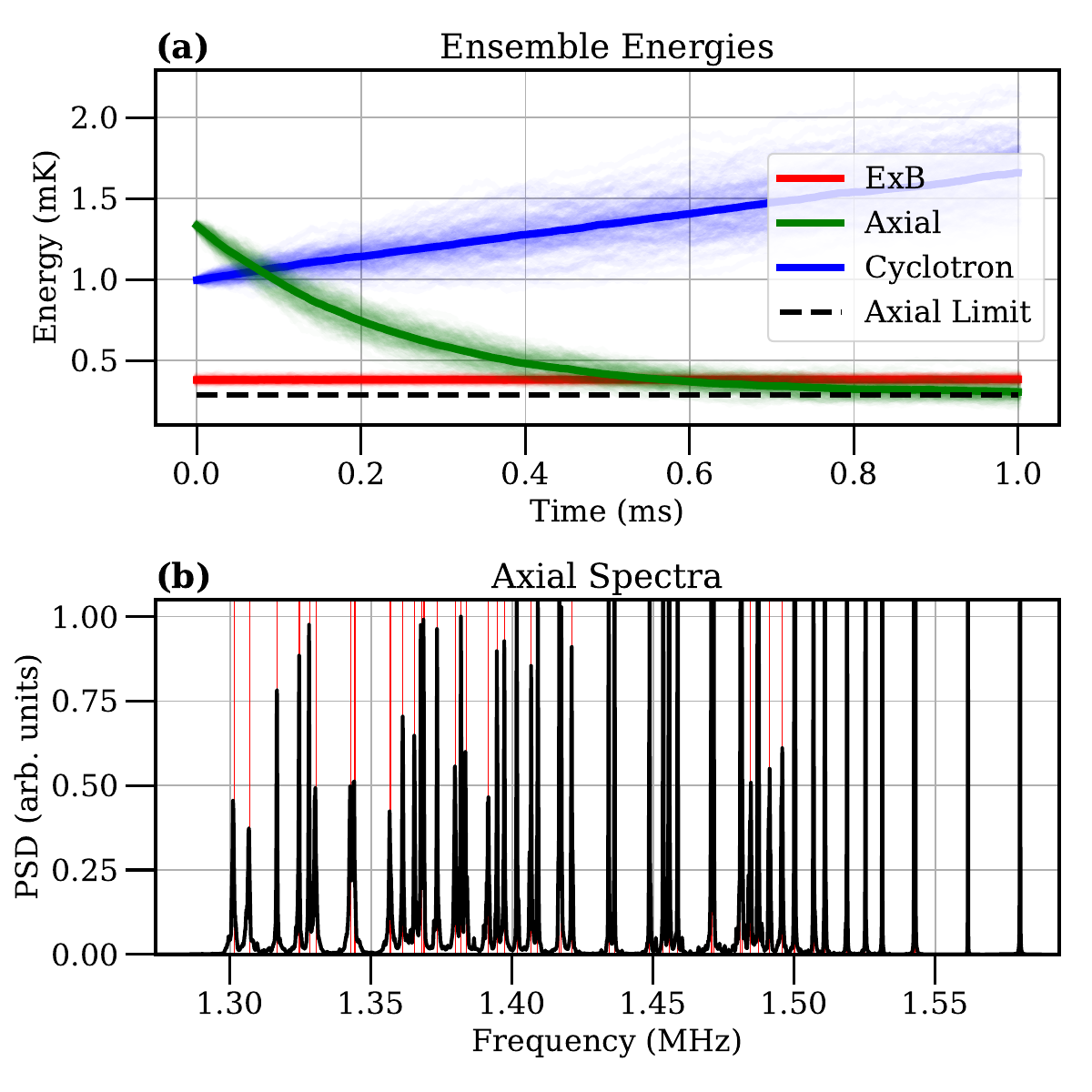}
\caption{\label{fig:laserCoolingSpectra}
\textbf{Laser cooling of drumhead modes following adiabatic $\exb$ cooling.}
Laser cooling is applied to the ensemble from Fig.~\ref{fig:adiabaticCooling} after the adiabatic ramp. 
Axial cooling beams are applied for 1 ms with fixed $\omega_r = 2\pi \times 180$ kHz.
\textbf{(a)} Mode-branch energies during laser cooling. 
The drumhead (axial) modes cool rapidly toward the Doppler limit, while the $\exb$ modes remain cold and the cyclotron modes absorb recoil heating.
\textbf{(b)} Power spectral density of the drumhead modes after 12.5 ms of free evolution following the axial cooling in (a). 
The ensemble-averaged spectrum (black) shows well-resolved peaks aligned with the theoretical mode frequencies (red lines), indicating high spectral resolution.
Thus, after preparation at $1~\mathrm{mK}$ with $\omega_r/2\pi=200~\mathrm{kHz}$ and an adiabatic ramp to $\omega_r/2\pi=180~\mathrm{kHz}$, laser re-cooling of the drumhead modes yields well-resolved drumhead spectra at the target operating point, preserving spectral resolution (cf. Ref.~\cite{Johnson2024}, Fig.~2(e)).
}
\end{figure}

\begin{figure}
\includegraphics[width=\linewidth]{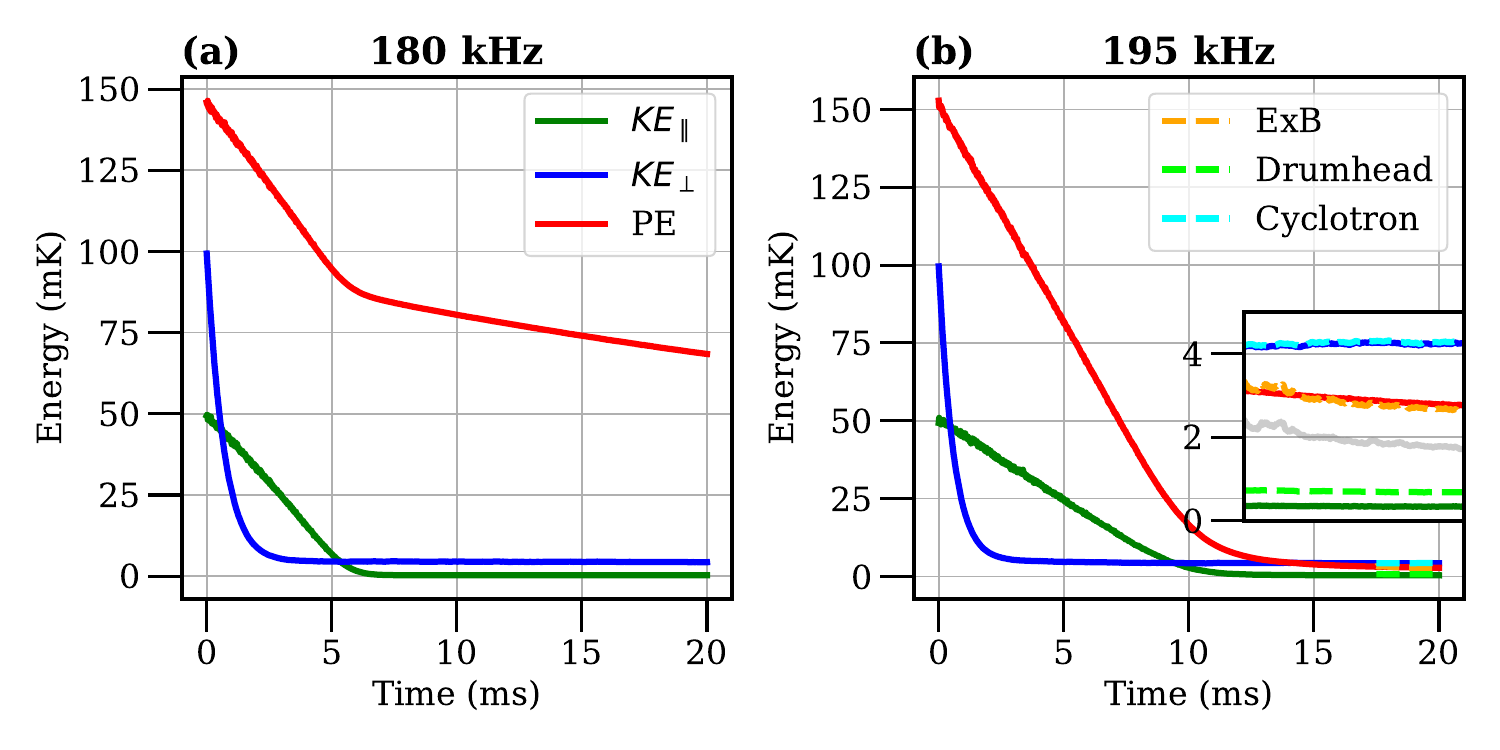}
\caption{\label{fig:coolingThermalEnsemble}
\textbf{Laser cooling of thermal ensembles near the planar-to-3D transition.}
An ensemble of 128 ion crystals with $N = 100$ ions is initialized at 100 mK using the Metropolis-Hastings algorithm. 
The crystals are initialized at $\omega_r = 2\pi \times 180 \text{ kHz and } 2\pi \times 195 \text{ kHz}$, with $\delta/\beta = 0.5$. 
Both ensembles are cooled for 20 ms using perpendicular and axial laser beams.
\textbf{(a)} Evolution of parallel and perpendicular kinetic energy ($\text{KE}_\parallel$, $\text{KE}_\perp$) and total potential energy (PE) for the 180 kHz ensemble. 
(Here, $\parallel$ denotes motion along the magnetic field $\mathbf{B}\parallel\hat z$ and $\perp$ denotes motion transverse to $\mathbf{B}$; $\text{KE}_\parallel=\sum_i \tfrac{1}{2} m_i \dot z_i^2$ and $\text{KE}_\perp=\sum_i \tfrac{1}{2} m_i (\dot x_i^2+\dot y_i^2)$.)
The system remains disordered and does not crystallize.
\textbf{(b)} Same for the 195 kHz ensemble. 
Strong nonlinear coupling between motional branches enables rapid energy exchange and cooling. 
After $\sim$17.5 ms, the system crystallizes, and mode energies can be computed.
\textbf{Inset:} Mode branch averages for the crystallized 195 kHz ensemble: $\exb$ modes are cooled to $\sim$2 mK, drumhead modes to $\sim$1 mK, and cyclotron modes to $\sim$4 mK.
The $y$-axis is the average energy in millikelvin (mK) units. 
The $x$-axis is the last 2.5 ms of the evolution.
The solid gray line represents the $\exb$ mode branch average without the inclusion of the COM mode, which is not coupled to the other modes and remains hot.
}  
\end{figure}

\begin{figure}
\includegraphics[width=\linewidth]{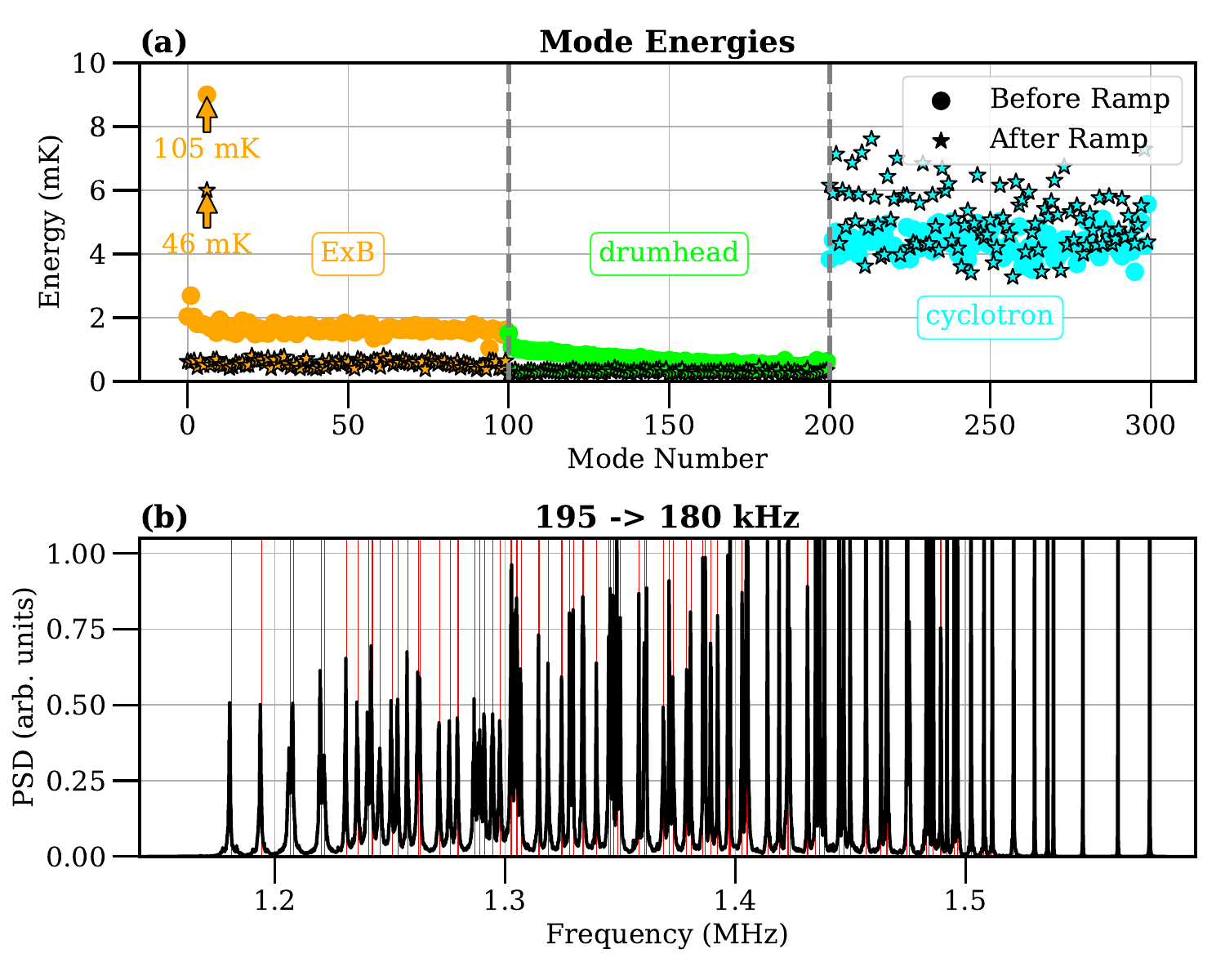}
\caption{\label{fig:mode_energies_and_spectra}
\textbf{Final cooling and spectral resolution after adiabatic ramp.}
The 195 kHz thermal ensemble from Fig.~\ref{fig:coolingThermalEnsemble}(b) is ramped to 180 kHz over 20 ms using a half-cosine ramp. 
After the ramp, axial laser cooling is applied for 1 ms to further cool the drumhead modes.
\textbf{(a)} Mode energies of the most common crystal configuration, before (circles) and after (stars) the ramp and axial re-cooling. 
All $\exb$ modes are cooled to below 1 mK, except the center-of-mass (COM) mode, which remains hot due to lack of coupling to any other modes.
\textbf{(b)} Power spectrum of the drumhead modes after 12.5 ms of free evolution. 
Peaks align with theoretical mode frequencies (vertical lines), indicating high spectral resolution.
This result demonstrates the effectiveness of the full cooling protocol---nonlinear coupling, adiabatic ramping, and final laser cooling---for preparing well-resolved planar ion crystals from thermal ensembles.
}
\end{figure}

To evaluate how the ramp affects the ion crystal, we simulate three different rotating wall ramp protocols: a linear ramp with $\delta$ held constant while $\omega_r$ and hence $\beta$ is lowered, a linear ramp with fixed $\delta/\beta$, and a half-cosine ramp with fixed $\delta/\beta$, shown in Fig.~\ref{fig:compareRamps}.  
In all cases, an $N = 50$ ion crystal is initialized in the minimum potential energy configuration at $\omega_r = 2 \pi \times 200$ kHz. 
The rotating wall frequency is then ramped down to $\omega_r = 2 \pi \times 190$ kHz with the ramp time $T = 1$ ms.
Unless otherwise noted, all energies are expressed in units of millikelvin (mK), defined as $E/(k_B)\times 10^3$, where $k_B$ is the Boltzmann constant.
Details of all simulation parameters and initialization procedures are provided in Appendix~\ref{app:simulations}. 

In Fig.~\ref{fig:compareRamps}(a), the linear rotating wall frequency ramp is applied with constant $\delta$. 
As the rotation frequency $\omega_r$ is decreased, the planar confinement parameter $\beta$ is also decreased, resulting in an increased ratio of $\delta/\beta$.
This causes the ion crystal to elongate in the $y$ direction, as the ions reorganize to a new equilibrium configuration, leading to heating of the ion crystal during the ramp.

In Fig.~\ref{fig:compareRamps}(b), the same linear ramp is applied with the ratio of $\delta/\beta$ fixed.
This prevents the ion crystal from reorganizing during the ramp; however, the impulsive Euler forces from the sudden start and stop of the ramp excite the rocking motion of the ion crystal.
Although subtle on the scale of the ion trajectories, the collective motion inferred from the time evolution encoded in Fig.~\ref{fig:compareRamps}(b) indicates that the rocking mode is excited during the ramp, leading to heating of the ion crystal.

In Fig.~\ref{fig:compareRamps}(c) a half-cosine ramp is applied with fixed $\delta/\beta$, while Fig.~\ref{fig:compareRamps}(b) shows an equal-duration linear ramp.
The half-cosine ramp has zero-slope endpoints and thus avoids impulsive Euler forces, but with a short duration $T=1~\mathrm{ms}\sim 2\pi/\omega_{\min}$ the motion is not fully adiabatic.
To make the geometric relation clear, light-gray segments in Fig.~\ref{fig:compareRamps}(b,c) connect each ion to the origin, showing that initial and final positions are collinear with the origin and hence related by a global rescaling with preserved aspect ratio.
Meanwhile, solid black segments connect each ion’s initial equilibrium site to its final rescaled equilibrium site predicted by Eq.~(\ref{eq:rescaled_eq}) (Appendix~\ref{app:adiabaticEnergy}), depicting the instantaneous-equilibrium locus during the ramp.
Because $T$ is short, ions deviate transiently from these loci: the equilibrium radius changes under the global rescaling, and the Euler force from $\dot{\omega}_r(t)$ drives azimuthal displacements.
By $t=T$ the half-cosine ramp’s boundary conditions return the ions to the rescaled equilibrium and at rest in the rotating frame, whereas the linear ramp leaves residual azimuthal rocking.
All other ramps used in the paper satisfy $T\gg 2\pi/\omega_{\min}$ and are adiabatic.

To remain in the adiabatic regime, the mode frequencies must be varied slowly compared to the period of the mode.
This presents a challenge for the low-frequency $\exb$ modes, which impose long ramp times---unless their frequencies are increased.
One way to do this is to increase the strength of the rotating wall, $\delta$. 

Figure~\ref{fig:varyWallStrength}(a) shows the lowest frequency $\exb$ mode as calculated by a linear analysis \cite{Shankar2020,Wang2013}, versus the rotating wall strength, $\delta$, for $N =$ 20, 30, 40, and 50 ion crystals.
The analysis is performed in the minimum potential energy configuration at $\omega_r = 2 \pi \times 200$ kHz with different rotating wall strengths, $\delta/\beta$ ranges from 0.1 to 0.5.
As the strength of the rotating wall is increased, the lowest frequency $\exb$ mode increases significantly in frequency, improving the dynamics of the ramp.
This behavior is reminiscent of structural instabilities in RF-trapped planar ion crystals, where increasing anisotropy stabilizes the crystal and raises the melting threshold, while larger systems exhibit lower-frequency modes and melt more readily \cite{Pashinsky2025}.

In Fig.~\ref{fig:varyWallStrength}(b), the potential energy difference from the minimum potential energy configuration of the ion crystal averaged over 1 ms after the half-cosine ramp is plotted versus the ratio $\delta/\beta$.
Since the $\exb$ modes contribute the majority of the potential energy of the ion crystal, this quantity is a good measure of the heating of these modes during the ramp.
The ion crystals are ramped down to $\omega_r = 2 \pi \times 180$ kHz with $T = 1$ ms.
For values of $\delta/\beta \gtrsim 0.3$, the potential energy of the ion crystal is near zero after the ramp, suggesting that the ramp produces minimal heating and the ion crystal is near the minimum potential energy configuration at the new rotation frequency.  
However, for smaller values of $\delta/\beta$, the ion crystal is heated during the ramp, as the potential energy of the ion crystal is significantly above zero after the ramp.    
This apparent heating is worse for smaller values of $\delta/\beta$. 

\subsection{Demonstrating Adiabatic Cooling in Ideal Crystals}

To test the predictions of the adiabatic cooling theory, we simulated an ensemble of 128 ion crystals, each with $N = 54$ ions and $\delta/\beta=1/3$. 
All modes were initialized at 1 mK with random phases. 
Each ion crystal is evolved with a half-cosine ramp, from $\omega_i = 2\pi \times 200$ kHz to $\omega_f = 2\pi \times 180$ kHz over $T = 20$ ms.
For these parameters, the one-to-two plane transition occurs near $\omega_r = 2\pi \times 204$ kHz, so the system remains well below this threshold throughout the ramp, and nonlinear coupling between the drumhead and $\exb$ mode branches is weak.
This ramp decreases the planar confinement parameter from $\beta_i \approx 0.090$ to $\beta_f \approx 0.034$, a reduction of roughly 60\%.
According to (\ref{eqn:exbAdiabaticEnergy}), the energy of the $\exb$ modes is predicted to be reduced to roughly 0.4 mK after the ramp.
After the ramp, the ion crystals were evolved at constant $\omega_r$ for 1 ms, and the time-averaged energies of the modes were calculated.

In Fig.~\ref{fig:adiabaticCooling}(a) the time-averaged $\exb$ mode energies are plotted versus mode number.
Individual ion crystal energies are shown as small gray dots (128 per mode), and the ensemble averages are shown as red dots.
The theoretical predictions for each mode are plotted as red stars, confirming the adiabatic theory predictions.
In contrast to the $\exb$ modes, the drumhead modes increase in frequency during the ramp, resulting in heating of the modes, seen in Fig.~\ref{fig:adiabaticCooling}(b).   

While the ensemble-averaged $\exb$ mode energies agree well with the theoretical adiabatic predictions, there is a thermal spread in the mode energies across the ensemble of ion crystals. 
This spread is due to nonlinear coupling between the modes, which enables thermalization of the mode energies during the ramp and subsequent averaging period. 
Notably, COM modes in each branch (mode numbers 5 and 107) exhibit minimal variation across the ensemble, as the COM modes are not coupled to the other modes, and therefore do not participate in thermalization. 

The ramp used in Fig.~\ref{fig:adiabaticCooling} cools the $\exb$ modes to roughly 0.4 mK; however, the drumhead modes, especially the lower frequency ones---which increase significantly in frequency during the ramp---are heated to roughly 2 mK. 
One solution is to laser cool the drumhead modes after the ramp. 
Fortunately, because planar recoil from axial cooling beams is a \emph{momentum} impulse, it couples predominantly to the $\mathrm{KE}_\perp$-dominated cyclotron modes ($R_n \ll 1$), while the slow $\exb$ branch is $\mathrm{PE}$-dominated ($R_n \gg 1$) and couples only weakly to momentum kicks; thus most planar recoil is taken up by the cyclotron modes~\cite{Shankar2020}.
As a result, the $\exb$ modes will not be significantly heated by laser cooling of the drumhead modes. 
This is advantageous because heating of the cyclotron modes does not result in a broadening of the drumhead mode spectrum, whereas the difficult-to-cool $\exb$ modes do \cite{Shankar2020}.   
Therefore, we evolve the ensemble from Fig.~\ref{fig:adiabaticCooling} with axial cooling beams applied for 1 ms after the ramp.
In Fig.~\ref{fig:laserCoolingSpectra}(a), the energies of the cyclotron modes are slightly heated due to recoil; meanwhile, the drumhead motion is rapidly cooled to near the Doppler limit, and the $\exb$ modes remain essentially unchanged after the cooling process.    

To assess the impact of the ramp and cooling protocol on the spectral resolution of the drumhead modes, we calculate the power spectral density (PSD). 
Here “PSD” denotes the \emph{axial} (drumhead) power spectral density: from the time series $\{z_i(t),\dot z_i(t)\}_i$, we compute the discrete Fourier transforms $Z_i(\omega)$ and $\dot Z_i(\omega)$, form the (single--sided) power spectra $|Z_i(\omega)|^2$ and $|\dot Z_i(\omega)|^2$, sum these over ions and over the two series, and ensemble--average; planar $(x,y)$ signals are not included~\cite{Tang2019}.
The PSD is calculated by evolving the ensemble for an additional 12.5 ms of free evolution after the axial cooling process.
Figure~\ref{fig:laserCoolingSpectra}(b) shows the average PSD of the drumhead modes across the ensemble, where the red vertical lines represent the theoretical predictions of the drumhead mode frequencies.   
The overlap of the red vertical lines with the peaks of the PSD indicates that the drumhead mode frequencies are well resolved.

Figure~2(f) of Ref.~\cite{Johnson2024} shows a resolved drumhead spectrum at $\omega_r/2\pi=204~\mathrm{kHz}$, where the planar ($\exb$) modes are coupled to the drumhead modes, while Fig.~2(e) shows an unresolved spectrum at $\omega_r/2\pi=180~\mathrm{kHz}$, where direct $\exb$ cooling is inefficient.
Here we initialize the $\exb$ branch to $1~\mathrm{mK}$ at $\omega_r/2\pi=200~\mathrm{kHz}$, close to the conditions of Fig.~2(f) of Ref.~\cite{Johnson2024}, and then adiabatically ramp to $\omega_r/2\pi=180~\mathrm{kHz}$ and re-cool the drumhead modes to demonstrate that the resolved drumhead spectrum is preserved at the lower operating frequency.
This motivates a practical sequence: cool at higher $\omega_r$ and then adiabatically ramp down.

\subsection{Nonlinear Coupling Enables Cooling from Disordered States}

So far, we have investigated cooling protocols that start at low temperatures, where the ions form a crystalline state.   
Next, in Fig.~\ref{fig:coolingThermalEnsemble}, we investigate laser cooling starting from experimentally relevant higher temperatures.  
Two thermal ensembles of 128 instances of 100-ion crystals are initialized at 100 mK with different rotation frequencies of $\omega_r = 2\pi\times 180$ kHz, and $\omega_r = 2\pi \times 195$ kHz.  
For $N = 100$ ions the rotation frequency 195 kHz is near the one-to-two plane transition, where the nonlinear coupling between the $\exb$ and drumhead modes is the strongest.  
At 100 mK, the ions are in a disordered state, and hence the mode energies cannot be calculated until sufficient cooling has occurred.  
After cooling, multiple ion crystal configurations are possible; however, for $\delta/\beta=0.5$ we find that one configuration occurs more frequently than the others.  
We note that this situation has been observed in experiments with planar ion crystals in rf traps~\cite{Kiesenhofer2023}.  
Both ensembles are cooled with planar and axial laser cooling beams for 20 ms, parameters are given in Table~\ref{table:simulation_params} in Appendix~\ref{app:simulations}. 
Throughout, we define “parallel” ($\parallel$) as along the magnetic field
$\mathbf{B}=B\hat{z}$ and “perpendicular” ($\perp$) as transverse to $\mathbf{B}$;
explicitly,
$\text{KE}_\parallel \equiv \sum_i \tfrac{1}{2} m\,\dot z_i^2$
and
$\text{KE}_\perp \equiv \sum_i \tfrac{1}{2} m\,(\dot x_i^2 + \dot y_i^2)$.

Figure~\ref{fig:coolingThermalEnsemble}(a) and (b) show the ensemble averages of the parallel and perpendicular kinetic energies $\text{KE}_\parallel\text{ and }\text{KE}_\perp$ and the total potential energy, $\text{PE}$, versus time. 
In both cases, $\text{KE}_\perp$ cools rapidly, indicating that the cyclotron motion of the ions is cooled most efficiently. However, the cooling of $\text{KE}_\parallel$ and $\text{PE}$ follow different trajectories between the two ensembles.  
For the slow $\exb$ branch, the motion is potential energy dominated, so low $\text{KE}_\perp$ can coexist with appreciable PE; in our notation this is $R_n \approx 500 \gg 1$ for the relevant modes~\cite{Shankar2020}. 
In the 180 kHz ensemble, [Fig.~\ref{fig:coolingThermalEnsemble}(a)], $\text{KE}_\parallel$ and $\text{PE}$ cool in tandem, indicating that energy is being exchanged between planar and axial motions of the ions; however, this ceases after roughly 5 ms---at which point the cooling of $\text{PE}$ slows. 
This is in contrast to the 195 kHz ensemble [Fig.~\ref{fig:coolingThermalEnsemble}(b)], where $\text{KE}_\parallel$ and $\text{PE}$ cool together for nearly the entire 20 ms.

After 20 ms, all the clouds of ions in the 195 kHz ensemble crystalize, allowing for the calculation of the mode energies, shown in the inset of Fig.~\ref{fig:coolingThermalEnsemble}(b). 
The inset shows that the $\exb$ modes are cooled to roughly 2 mK, the drumhead modes are cooled to the Doppler limit ($<0.5$ mK), and the cyclotron modes are cooled to roughly 4 mK. 
The $\exb$ mode branch average energy appears to have not equilibrated, suggesting that further cooling is possible with a longer cooling time.
This result demonstrates that operating near the one-to-two-plane transition enables efficient cooling from a disordered thermal state, validating the role of nonlinear coupling in the cooling process \cite{Johnson2024}.  

We note that the $\exb$ mode branch average energy includes the COM mode, which remains at an elevated temperature after the cooling process, as it is not coupled to any other modes.   
Removing the COM mode from the average shifts the energy of the $\exb$ modes down to roughly 2 mK, as shown in the inset of Fig.~\ref{fig:coolingThermalEnsemble}(b).    

\subsection{Combining Techniques: Ramping After Nonlinear Cooling}

We now combine the insights gained so far by first laser cooling close to the one-to-two plane transition frequency at 195 kHz and then adiabatically ramping the rotation frequency down to 180 kHz. 
Figure~\ref{fig:mode_energies_and_spectra}(a) shows the average mode energies of the most common ion crystal configuration in the 195 kHz ensemble after nonlinear coupling (dots) and after the ramp to 180 kHz and axial re-cooling (stars).   
Except for the COM mode, all of the $\exb$ modes are cooled to sub-millikelvin temperatures.
The COM mode does not couple nonlinearly to any other modes due to the translational invariance of the Coulomb interaction and the harmonic form of the trapping potential, and thus can only be cooled via direct interaction with the perpendicular laser beam. 
In real experiments, this symmetry is broken by imperfections such as trap anharmonicities and the accumulation of impurity ions~\cite{McAneny2013,Jensen2004,Sawyer2015,Dubin2013}, which will introduce weak coupling channels for cooling the COM mode.
Moreover, experimental protocols typically involve many repeated sequences with recooling applied between shots. 
Over time, this allows the COM mode to reach a lower steady-state temperature. 
While the single-shot simulations presented here retain a hot COM mode as a consequence of the idealized model, this behavior is expected, and importantly, has little practical consequence. 
Because the COM mode remains decoupled from the rest of the system, its elevated energy does not degrade the resolution of the drumhead mode spectra, nor does it contribute to additional heating.

In Fig.~\ref{fig:mode_energies_and_spectra}(b), the ensemble is evolved for 12.5 ms of free evolution, and the PSD of the drumhead modes is calculated for the most common ion crystal configuration in the ensemble (using same procedure as Fig.~\ref{fig:laserCoolingSpectra}).  
Remarkably, the drumhead mode spectra is resolved well, even though the ion crystals are twice as large as the ones in Fig.~\ref{fig:laserCoolingSpectra}(b) and initialized at a much higher temperature.  

These results confirm that strong nonlinear coupling near the one-to-two-plane transition enables efficient cooling from disordered thermal states. 
The adiabatic ramp maintains spectral resolution without reheating. 
Together, these techniques offer a practical path for preparing cold, well-resolved planar ion crystals from high-temperature initial conditions.

\section{Conclusion and Outlook}

We have shown that nonlinear coupling between the $\exb$ and drumhead modes can be dynamically tuned by adiabatically varying the rotation frequency of the ion crystal during experiments. 
This enables the preparation of crystals with all motional modes cooled to millikelvin temperatures at experimentally relevant rotation frequencies—setting the stage for improved spectral resolution and enhanced quantum information processing.

Unlike traditional laser cooling techniques, this method requires no additional laser beams or hardware changes. 
Instead, it exploits the intrinsic motional dynamics of the ion crystal, controlled through simple frequency ramps.

Experimental implementation will require precise control of the rotating wall potential over millisecond timescales, but the technique remains compatible with existing Penning trap systems. 
Future work will focus on optimizing ramping protocols, for example, by investigating the impact of different rotating wall geometries, such as the triangular rotating wall potential used in Ref.~\cite{McMahon2022}, which may lead to more stable ion crystal configurations \cite{Dubin2013,Khan2015}, allowing for faster ramps. 
These results open a path toward more reliable quantum control and invite further exploration of nonlinear dynamics as a tool for state preparation in large-scale ion crystals.

\begin{acknowledgments}
This work was supported by the U.S. Department of Energy (W.J., J.Z., S.E.P., grant number DE-SC0020393).
 A.S. acknowledges support through a New Faculty Initiation Grant (NFIG) from IIT Madras. Support is also acknowledged from the U.S. Department of Energy, Office of Science, NQIS QSA Research Center, National Institute of Standards and Technology, and AFOSR grant FA9550-25-1-0080 (J.J.B., B.B.). 

\end{acknowledgments}

\appendix

\section{\label{app:normalModes} Canonical Mode Variables of Ion Crystals} 

The normal modes of the ion crystal can be calculated in the position and velocity coordinates of the ions \cite{Shankar2020}.
The linearized energy of the system can be expressed as a matrix equation of the form: 
\begin{equation}
    E = \frac{1}{2} \mathbf{Z_\text{v}}^T \cdot \mathbb{E} \cdot \mathbf{Z_\text{v}}, \quad \mathbb{E} = \begin{pmatrix}
    \mathbb{K} & \mathbb{0} \\
    \mathbb{0} & \mathbb{M}
    \end{pmatrix}, \quad \mathbf{Z_\text{v}} = \begin{pmatrix}
    \mathbf{q} \\
    \dot{\mathbf{q}}
    \end{pmatrix} 
\label{eqn:energyMatrix}
\end{equation}
where $\mathbf{q} = (\mathbf{x}, \mathbf{y}, \mathbf{z})$ is the vector of the ion positions, with $\mathbf{x} = (x_1, x_2 \dots x_N)$ and $\mathbf{y},\mathbf{z}$ ordered similarly; meanwhile, $\dot{\mathbf{q}} = (\dot{\mathbf{x}}, \dot{\mathbf{y}}, \dot{\mathbf{z}})$ is the vector of the ion velocities. 
$\mathbb{K}$ is the stiffness matrix and $\mathbb{M}$ is the mass matrix.   
The relationship between the energy matrix $\mathbb{E}$ and the normal modes of the system is not straightforward; however, by transforming the coordinates of the system to the generalized canonical position and momentum variables of the system, the Hamiltonian matrix can be constructed, from which the normal modes can be calculated simply.    
The transformation from velocity to canonical coordinates, $\mathbf{Z_\text{v}} \rightarrow \mathbf{Z_\text{c}}$, $\mathbf{Z_\text{c}} = \mathbb{T} \cdot \mathbf{Z_\text{v}}$, is given by the transformation matrix $\mathbb{T}$: 
\begin{equation}
    \mathbb{T} = \begin{pmatrix}
    \mathbb{I} & \mathbb{0} \\
    \mathbb{B} & \mathbb{M} \end{pmatrix}, 
    \quad \mathbb{B} = \begin{pmatrix}
        \mathbb{0} & \mathbb{C} & \mathbb{0} \\
        -\mathbb{C} & \mathbb{0} & \mathbb{0} \\
        \mathbb{0} & \mathbb{0} & \mathbb{0} 
    \end{pmatrix}, 
\label{eqn:transformationMatrix}
\end{equation}
where $\mathbb{I}$ is the identity matrix, $\mathbb{0}$ is the zero matrix, and $\mathbb{C}$ is the matrix with coefficients $\frac{1}{2} m ( \omega_c - 2 \omega_r)$ along its diagonal.
Using the transformation matrix, the energy matrix can be expressed in the canonical coordinates as:    
\begin{equation}
    \mathbb{H} = (\mathbb{T}^{-1})^T \cdot \mathbb{E} \cdot \mathbb{T}^{-1}, 
\label{eqn:energyMatrixCanonical}
\end{equation}
where $\mathbb{H}$ is the Hamiltonian matrix in the canonical coordinates, and $(\mathbb{T}^{-1})^T$ is the transpose of the inverse of the transformation matrix.  
From here, the normal modes can be calculated by diagonalizing the Hamiltonian matrix \cite{Dubin2020}.
Each normal mode can then be expressed as a harmonic oscillator Hamiltonian, Eq.~(\ref{eqn:mode_hamiltonian}).

\section{\label{app:adiabaticEnergy} Derivation of Adiabatic Energy Scaling} 

To support the energy-scaling model used in Section~\ref{sec:adiabaticCooling}, we present a detailed derivation of the adiabatic energy scaling for the $\exb$ modes.
This treatment also addresses a potential criticism: that lowering the planar mode energy is simply a consequence of reducing confinement. 
We show that while this is true, it does not compromise the validity of our approach, as the system remains in the weakly nonlinear regime throughout.
Remaining in the weakly nonlinear regime ensures that the drumhead mode spectra are not significantly broadened by in-plane motions, and that the normal mode picture remains valid.    

\subsection*{Guiding Center Approximation and $\exb$ Mode Scaling}

In the limit where the cyclotron frequency $\omega_c$ is much larger than the planar mode frequencies, the equations of the planar motion reduce to the guiding center approximation \cite{Dubin1988}. 
In matrix form, the linearized equations of motion for the $\exb$ motion of the ions can be expressed as:
\begin{equation}
    \dot{\mathbf{q}}= \frac{1}{m (\omega_c - 2\omega_r)} \mathbb{J} \cdot \mathbb{K} \cdot \mathbf{q}, 
\label{eqn:guidingCenter}
\end{equation}
where $\mathbb{K}$ is the planar stiffness matrix and $\mathbb{J}$ is an anti-symmetric matrix arising from the Lorentz force: 
\begin{equation}
    \mathbb{J} = \begin{pmatrix}
    \mathbb{0} & \mathbb{I} \\
    -\mathbb{I} & \mathbb{0}
    \end{pmatrix},
\label{eqn:matrixJ} 
\end{equation}
where \(\mathbb{I}\) is the identity matrix and \(\mathbb{0}\) is the zero matrix.

The eigenvalues of Eq.~(\ref{eqn:guidingCenter}) approximate the $\exb$ mode frequencies well in the large-cyclotron-frequency limit, appropriate for NIST experiments \cite{Bohnet2016}.
As we show below, these scale linearly with the planar confinement parameter $\beta$.

Restricting our attention to the motion in the crystal plane by setting $z_i = 0$, the potential energy is given by:
\begin{equation}
\begin{split}
    U\left(\{\mathbf{R}_i\}_{i=1}^N\right) = \frac{1}{2} \sum_{i=1}^{N}\sum_{j \ne i} \frac{k_e q^2}{|\mathbf{R}_i - \mathbf{R}_j|} \\
    + \frac{1}{2} m \omega_z^2 \beta \sum_{i=1}^N 
    [(1+\alpha)x_i^2 + (1-\alpha)y_i^2],
\end{split}
\end{equation}
where $\alpha = \delta / \beta $ is the anisotropy parameter, which is kept constant. 

The positions and energies of the ions can be rescaled into dimensionless form with characteristic length and energy scales:
\begin{equation} 
    l_0 = \left( \frac{k_e q^2}{1/2 \beta m \omega_z^2} \right)^{1/3}, \quad  E_0 = m \omega_z^2 l_0^2.
\end{equation}

Since $l_0 \propto \beta^{-1/3}$, the equilibrium positions of the ions scale as:
\begin{equation}
    R'_{\text{eq}} = \left( \frac{\beta}{\beta'} \right)^{1/3} R_{\text{eq}}.
    \label{eq:rescaled_eq}
\end{equation}

This implies that the Coulomb term, when differentiated twice to form the stiffness matrix, scales as $\sim 1/l_0^3 \propto \beta $. 
The trap term also scales with $\beta$, so overall:
\begin{equation}
    \mathbb{K}' = \left(\frac{\beta'}{\beta}\right) \mathbb{K}.
\end{equation}

Using the guiding center eigenvalue problem:
\begin{equation}
    \frac{1}{m(\omega_c - 2\omega_r)}\mathbb{J}\mathbb{K} |\mathbf{e}_n\rangle = -i \omega_n |\mathbf{e}_n\rangle,
\end{equation}
we find that all $\exb$ mode frequencies scale as:
\begin{equation}
    \omega_n' = \left( \frac{\beta'}{\beta} \right) \omega_n.
\end{equation}

Applying the adiabatic invariant $I_n = E_n / \omega_n $, we find Eq.~(\ref{eqn:exbAdiabaticEnergy}) of the main text. 

\subsection*{Nonlinearity Scaling and Validity of Approximation}

Although the \(\exb\) energy scales with confinement, one might worry that the reduction in confinement leads to a breakdown of the weakly nonlinear approximation.
To assess this, we define the nonlinearity parameter \(\epsilon\) as:
\begin{equation}
    \epsilon = \frac{q}{l_0},
\end{equation}
where \(q\) is the RMS displacement of ions from equilibrium.
If the displacements are small compared to the characteristic length scale of the system, then the system is in the weakly nonlinear regime, and the normal mode approximation is valid.

Using conservation of the action \( I \), we relate displacement to \(\beta\):
\begin{equation}
    q_n' = \sqrt{\frac{\beta}{\beta'}} q_n.
\end{equation}

Combining with \( l_0 \propto \beta^{-1/3} \), the nonlinearity parameter scales as:
\begin{equation}
    \epsilon' = \left( \frac{\beta}{\beta'} \right)^{1/6} \epsilon.
\end{equation}

This growth is weak: a 100-fold decrease in \( \beta \) leads to only a factor \( \sim 2.15 \) increase in nonlinearity. 
Thus, if the system starts in the weakly nonlinear regime, it remains so throughout the ramp, and the normal mode picture is valid. 
These results justify the use of a linear normal mode basis in modeling adiabatic cooling despite the decreased planar confinement.  

\section{\label{app:simulations} Simulation Parameters and Details} 

In this Section, we provide the simulation parameters and details used in each figure.
All simulations in this work use the following parameters, unless otherwise specified:
\begin{table}[H]
    \centering
    \begin{tabular}{l c}
    \hline\hline
    \textbf{Parameter} & \textbf{Value} \\[0.5ex]
    \hline
    Ion species & $^9$Be$^+$ \\[0.5ex]
    Magnetic field ($B$) & $4.4588$ T \\[0.5ex]
    Axial frequency ($\omega_z$) & $2\pi \times 1.58$ MHz \\[0.5ex]
    Time step ($dt$) & $1$ ns \\[0.5ex]
    Laser Wavelength ($\lambda$) & $313$ nm \\[0.5ex]   
    Cooling Linewidth ($\gamma_0$) & $2\pi \times 18$ MHz \\[0.5ex]
    Axial Beams Saturation ($S_\parallel$) & $5\times 10^{-3}$ \\[0.5ex]
    Axial Beams Detuning ($\Delta_\parallel$) & $ - \gamma_0/2$ \\[0.5ex]
    Planar Beam Saturation ($S_\perp$) & $1$ \\[0.5ex]
    Planar Beam Detuning ($\Delta_\perp$) & $ -\gamma_0/2$  \\[0.5ex] 
    Planar Beam Waist ($W_0$) & $30$ $\mu$m \\[0.5ex]
    Planar Beam Offset ($y_0$) & $15$ $\mu$m \\[0.5ex]  
    \hline\hline
    \end{tabular}
    \caption{Summary of simulation parameters used in numerical studies.}
    \label{table:simulation_params}
\end{table}
\noindent Meanwhile, the rotation frequency $\omega_r$ and the strength of the rotating wall $\delta$ were varied during simulations.
The temperature and procedure for initializing the ion crystals are also varied depending on the figure.

\textbf{Figure \ref{fig:compareRamps}:}
In these simulations, the frequency of the rotating wall varied according to the ramping protocol, and the strength of the rotating wall was either fixed, as in Fig.~\ref{fig:compareRamps}(a) or varied, as in Fig.~\ref{fig:compareRamps}(b) and (c) where $\delta/\beta = 0.5$ was kept constant. 
The equilibrium positions of the ions were calculated, and the ions were initialized in equilibrium at $\omega_r = 2 \pi \times 200$ kHz with $\delta/\beta = 0.5$.
The rotating wall frequency was ramped down to $\omega_r = 2 \pi \times 190$ kHz with $T = 1$ ms.  
The molecular dynamics simulations \cite{Tang2019} are performed in the lab frame, therefore, to plot the ion trajectories in the co-rotating frame, the coordinates of the ions are transformed. 
At each time step, the rotated coordinates are given by a rotation matrix, which depends only on the integrated rotation angle of the rotating wall. 
For the linear ramp, the integrated rotation angle is given by $\theta_r(t) = \omega_i t + \frac{1}{2} (\omega_f - \omega_i) t^2/T$, where $T$ is the ramp time, $\omega_i$ is the initial rotating wall frequency, and $\omega_f$ is the final rotating wall frequency.    
For the cosine ramp, the integrated rotation angle is given by $\theta_r(t) = \frac{1}{2} (\omega_i + \omega_f) t + \frac{1}{2} (\omega_i - \omega_f) \frac{T}{\pi} \sin(\frac{\pi t}{T})$. 
The rotation matrix is given by: 
\begin{equation}
    \mathbb{R(\theta_r)} = \begin{pmatrix}
        \cos(\theta_r) & -\sin(\theta_r) \\
        \sin(\theta_r) & \cos(\theta_r)
    \end{pmatrix}.
\end{equation}
When applied to the planar x-y coordinates of the ions, this rotates the coordinates into the co-rotating frame of the rotating wall with accumulated rotation angle $\theta_r$.    

\textbf{Figure \ref{fig:varyWallStrength}:}
In these simulations, all ion crystals were initialized in equilibrium at $\omega_r = 2 \pi \times 200$ kHz with different rotating wall strengths, corresponding values of $\delta/\beta = 0.1 - 0.5$.  
The rotating wall frequency was ramped down to $\omega_r = 2 \pi \times 180$ kHz with $T = 1$ ms.
After ramping, the ion crystals were evolved for 1 ms, and the time-averaged PE in the $\omega_r = 2\pi\times 180$ kHz frame is calculated.

\textbf{Figure \ref{fig:adiabaticCooling}:}
In these simulations, an ensemble of 128 different 54-ion crystals with $\delta/\beta = 1/3$ was initialized with all modes in each crystal given an amplitude corresponding to a temperature of 1 mK and a random phase. 
The mode initialization was described in detail in \cite{Tang2019,Tang2021}.
The value of $\delta/\beta$ was set to 1/3 and held constant during the ramp.
The half-cosine ramp was applied to the rotating wall frequency, $\omega_r$, from $2\pi\times200$ kHz to $2\pi\times180$ kHz with $T = 20$ ms.
After the ramp, the system was evolved at constant $\omega_r$ for 1 ms, and the time-averaged energies were calculated for each mode.
Next, the ensemble average for each mode is calculated.
These averages are plotted as large solid dots---this is possible because the ion crystals remained in the same configuration during the ramp. 

\textbf{Figure \ref{fig:laserCoolingSpectra}:}
The same ensemble in Fig.~\ref{fig:adiabaticCooling} was evolved with axial laser cooling beams applied for 1 ms after the ramp. 
The laser cooling parameters for the cooling beams are similar to those used in Ref.~\cite{Johnson2024} and are given in Table~\ref{table:simulation_params}.  
The 1 ms averaging time was not applied before the laser cooling process. 
The PSD was calculated by evolving the ensemble for an additional 12.5 ms of free evolution after the axial cooling process.
The positions and velocities of the ions were saved every 250 time steps, which allows the mode spectra to be resolved to 160 Hz. 
To calculate the power spectra, the lab frame coordinates, both velocity and position, were transformed into the rotating frame. 
The PSD was calculated using the Fourier transform of the time-series data, and the average PSD across the ensemble is plotted.

\textbf{Figure \ref{fig:coolingThermalEnsemble}:}
In these simulations, we initialized two ensembles of 128 distinct 100-ion crystals using the Metropolis–Hastings algorithm \cite{Pathria2011} at a temperature of 100 mK. 
The ensembles differ only in their rotation frequencies: $\omega_r = 2\pi \times 180$ kHz and $\omega_r = 2\pi \times195$ kHz.
For the ratio of $\delta/\beta$ = 0.5, and $N = 100$ ions, $\omega_r = 2\pi\times 195$ kHz is near the one-to-two plane transition, where nonlinear coupling between $\exb$ and drumhead modes is the strongest.  
Both ensembles were cooled with the laser cooling parameters and setup described in Ref.~\cite{Johnson2024}, given in Table~\ref{table:simulation_params}, for 20 ms.    

\begin{figure}
\includegraphics[width=\linewidth]{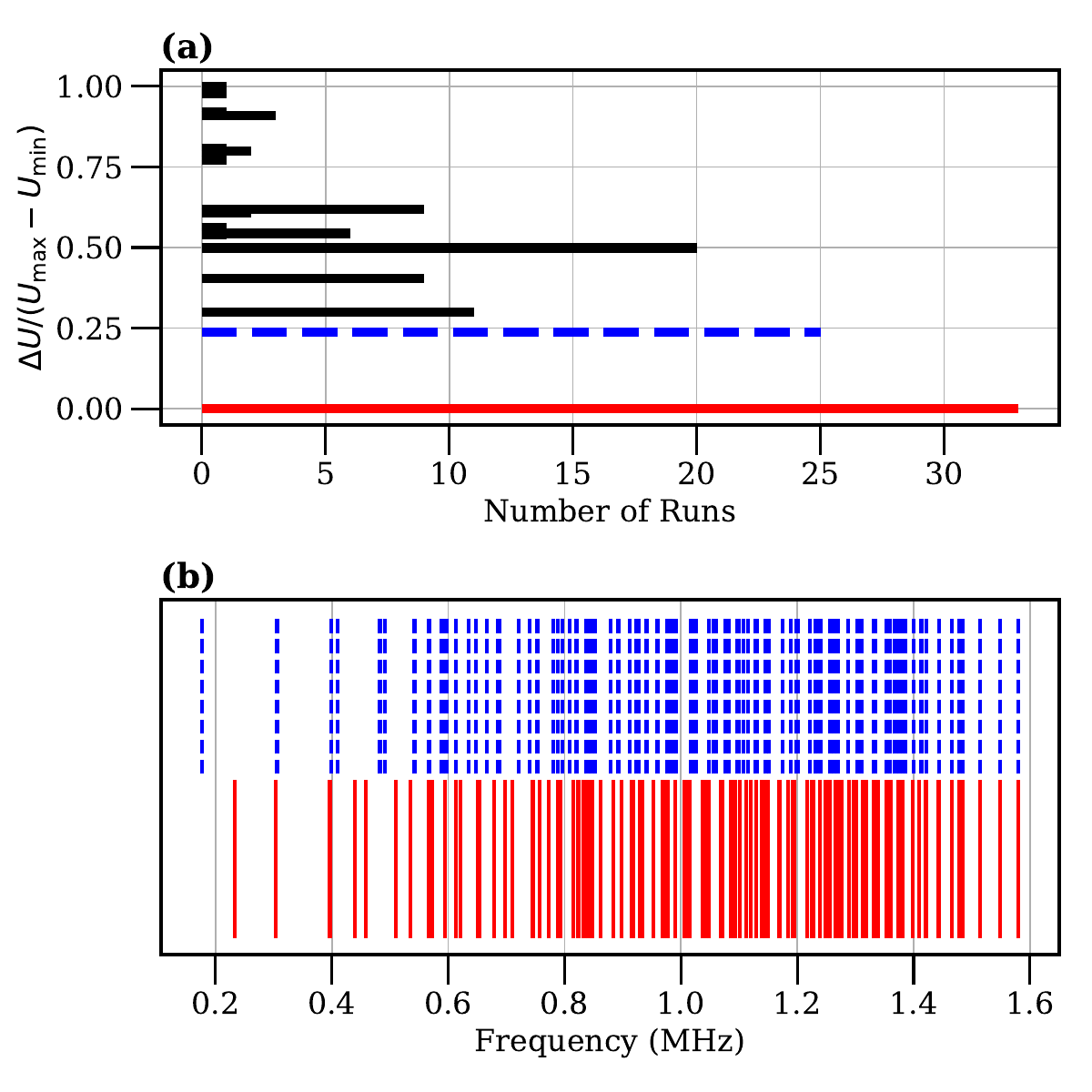}
\caption{\label{fig:configs_from_thermal_ensemble}
\textbf{Crystallization and mode structure of configurations in the 195 kHz ensemble.}
\textbf{(a)} Number of ion crystal configurations observed in the 195 kHz ensemble, plotted by potential energy relative to the lowest and highest energy configuration observed. 
Potential energies are computed using Eq.~(\ref{eqn:potentialEnergy}) and the numerically optimized equilibrium positions. 
The most common configuration (red line) corresponds to the global minimum. 
The second most common configuration (blue dashed line) is slightly higher in energy. 
All other configurations are shown in black.
\textbf{(b)} Drumhead mode frequencies for the most common (red solid) and second most common (blue dashed) configurations. 
Higher-order (lower-frequency) modes differ significantly due to their sensitivity to ion positions.
These results justify focusing on the most common configuration when analyzing mode energies and spectra after cooling.
}
\end{figure}

A difficulty arises when calculating the mode branch energies, as not every ion crystal in the ensemble had the same configuration at the end of the cooling process in Fig.~\ref{fig:coolingThermalEnsemble}(b).   
To circumvent this, we calculated the mode energies for the most common ion crystal configuration in the ensemble. 
The configuration was determined from the last frame of the simulation, which was used as a seed for a numerical optimization routine using Eq.~(\ref{eqn:potentialEnergy}) to find the equilibrium positions of the ions.
The mode spectra and energies were then calculated using the equilibrium found. 
For context, in Fig.~\ref{fig:configs_from_thermal_ensemble}(a), the potential energy of each configuration is plotted for the 195 kHz ensemble, where the most common configuration is indicated by the solid red line, and the second most common configuration is indicated by the dashed blue line. 
The most common configuration was used to calculate the mode energies and spectra in Fig.~\ref{fig:coolingThermalEnsemble} and Fig.~\ref{fig:mode_energies_and_spectra} of the main text. 
The most common configuration was also the one with the lowest potential energy. 
A similar situation arises in the experiment of Ref.~\cite{Kiesenhofer2023}, where a planar ion crystal was experimentally prepared in a radio-frequency trap in a number of nearly degenerate configurations.

After the ramp was applied, the mode analysis was performed on the most common equilibrium configuration in the ensemble. 
These configurations were determined from the last frame of the simulation at the new rotation frequency $\omega_r = 2\pi\times180$ kHz.
Due to the low-energy state of the $\exb$ modes, the ion crystals did not reorganize during the ramp, and the ion crystals remained in a configuration similar to the one before the rotating wall ramp was applied.
The spectra of the drumhead modes were calculated from free evolution of the system for 12.5 ms after the ramp. 
The spectra was plotted only for the most common ion-crystal configuration in the ensemble; however, the results are similar for the other configurations in the ensemble, with good agreement between the mode frequencies and the spectra.

\section{\label{app:MH} Metropolis--Hastings Thermal Initialization}

We generate thermal ensembles by sampling from the Boltzmann distribution of the full simulation Hamiltonian. 
The target distribution matches the molecular–dynamics (MD) Hamiltonian used in the main text so that the sampled states can be used directly as initial conditions for cooling ramps and adiabatic protocols.

\subsection*{Algorithm and settings}

We sample $(\mathbf{x},\mathbf{v})\in\mathbb{R}^{6N}$ from
\begin{equation}
\pi(\mathbf{x},\mathbf{v}) \propto \exp\!\left[-\frac{H(\mathbf{x},\mathbf{v})}{k_B T}\right],
\end{equation}
with
\begin{equation}
H(\mathbf{x},\mathbf{v}) = U_{\mathrm{trap}}(\mathbf{x}) + U_{\mathrm{Coul.}}(\mathbf{x}) + \tfrac{1}{2}\sum_{i=1}^{N} m\,\mathbf{v}_i^2.
\end{equation}
The potential $U_{\mathrm{trap}}+U_{\mathrm{Coul.}}$ is evaluated with the same routines and parameters used during MD; any constant reference offset in $U$ is irrelevant because it cancels in energy differences.

Each Metropolis--Hastings step selects one ion uniformly at random and proposes independent Gaussian increments for its position and velocity, $\delta\mathbf{x}_i,\delta\mathbf{v}_i \sim \mathcal{N}(\mathbf{0},\sigma^2 \mathbb{I}_3)$, yielding the trial state $\mathbf{x}_i'=\mathbf{x}_i+\delta\mathbf{x}_i$, $\mathbf{v}_i'=\mathbf{v}_i+\delta\mathbf{v}_i$, with all other ions unchanged. The proposal is symmetric, so the acceptance probability depends only on the energy change $\Delta H = H'-H$ via
\begin{equation}
\alpha = \min\left\{1,\exp\!\left[-\frac{\Delta H}{k_B T}\right]\right\}.
\end{equation}
We scale the proposal width with temperature, $\sigma \propto \sqrt{T}$ in code units, to maintain practical acceptance rates across the temperatures considered.

For ensembles near a crystalline configuration we initialize from the equilibrium positions (optionally adding small normal-mode displacements as a warm start). For higher temperatures (e.g., $100\,\mathrm{mK}$) MH initialization yields a disordered cloud without a normal–mode description, which may subsequently crystallize during the cooling protocol.
Unless otherwise noted we use $n_{\mathrm{burn}}=2\times 10^{4}$ burn–in steps to reach the desired $T$, thin by taking one sample every $n_{\mathrm{thin}}=10^{3}$ steps to avoid correlations between sampled states, and assemble $N_{\mathrm{ens}}=128$ samples per condition. 
We monitor the running mean energy per degree of freedom, $\bar{E}/(3N)$, to confirm stationarity and consistency with the target $T$ within expected fluctuations.

\bibliography{bib}{}
\bibliographystyle{apsrev4-2}

\end{document}